\documentclass[aps,english,prb,showpacs,preprint,superscriptaddress]{revtex4-1}
\usepackage{graphicx}

\begin{document}

\title
{Effects of charging and electric field on the properties of silicene and germanene}

\author{H. Hakan G\" urel}
\affiliation{UNAM-National Nanotechnology Research Center, Bilkent University, 06800 Ankara, Turkey}
\affiliation{Institute of Materials Science and Nanotechnology, Bilkent University, Ankara 06800, Turkey}
\address{ Tecnology Faculty, Department of Information Systems Engineering, Kocaeli University, Kocaeli 41380, Turkey}

\author{V. Ongun \"{O}z\c{c}elik}
\affiliation{UNAM-National Nanotechnology Research Center, Bilkent University, 06800 Ankara, Turkey}
\affiliation{Institute of Materials Science and Nanotechnology, Bilkent University, Ankara 06800, Turkey}

\author{S. Ciraci}\email{ciraci@fen.bilkent.edu.tr}
\affiliation{UNAM-National Nanotechnology Research Center, Bilkent University, 06800 Ankara, Turkey}
\affiliation{Institute of Materials Science and Nanotechnology, Bilkent University, Ankara 06800, Turkey}
\affiliation{Department of Physics, Bilkent University, Ankara 06800, Turkey}

\begin{abstract}

Using first-principles Density Functional Theory calculations, we showed that electronic and magnetic properties of bare and Ti adatom adsorbed single-layer silicene and germanene, which are charged or exerted by a perpendicular  electric field are modified to attain new functionalities. In particular, when exerted by a perpendicular electric field, the symmetry between the planes of buckled atoms is broken to open a gap at the Dirac points. The occupation of $3d$-orbitals of adsorbed Ti atom changes with charging or applied electric field to induce significant changes of magnetic moment. We predict that neutral silicene uniformly covered by Ti atoms becomes a half-metal at a specific value of coverage and hence allows the transport of electrons in one spin direction, but blocks the opposite direction.  These calculated properties, however exhibit a dependence on the size of the vacuum spacing between periodically repeating silicene and germanene layers, if they are treated using plane wave basis set within periodic boundary condition. We clarified the cause of this spurious dependence and show that it can be eliminated by the use of local orbital basis set.

\end{abstract}

\maketitle

When charged or subjected to an external electric field, the electronic structure of two-dimensional (2D) materials show important modifications. In particular graphene, a single layer of honeycomb structure of carbon atoms, is a semimetal and has an ambipolar character with $\pi$ and $\pi^*$ bands, which cross linearly at the Fermi level. Depending on the polarity of the excess charge in graphene, the Dirac points shift up or down the Fermi level. For example, the excess electrons cause the Dirac points to dip below the Fermi level. This way, semimetal graphene changes into a metal. When electrons are depleted, the Dirac points move up and graphene becomes hole-doped. Electric field applied perpendicular to graphene layers breaks the symmetry between both sides of the plane of carbon atoms. However, the $\pi$ and $\pi^*$ bands continue to cross. The effect of excess charge and perpendicular electric field on the electronic structure of graphene were investigated earlier\cite{topsakal1}.

\begin{figure}
\begin{center}
\includegraphics[width=12cm]{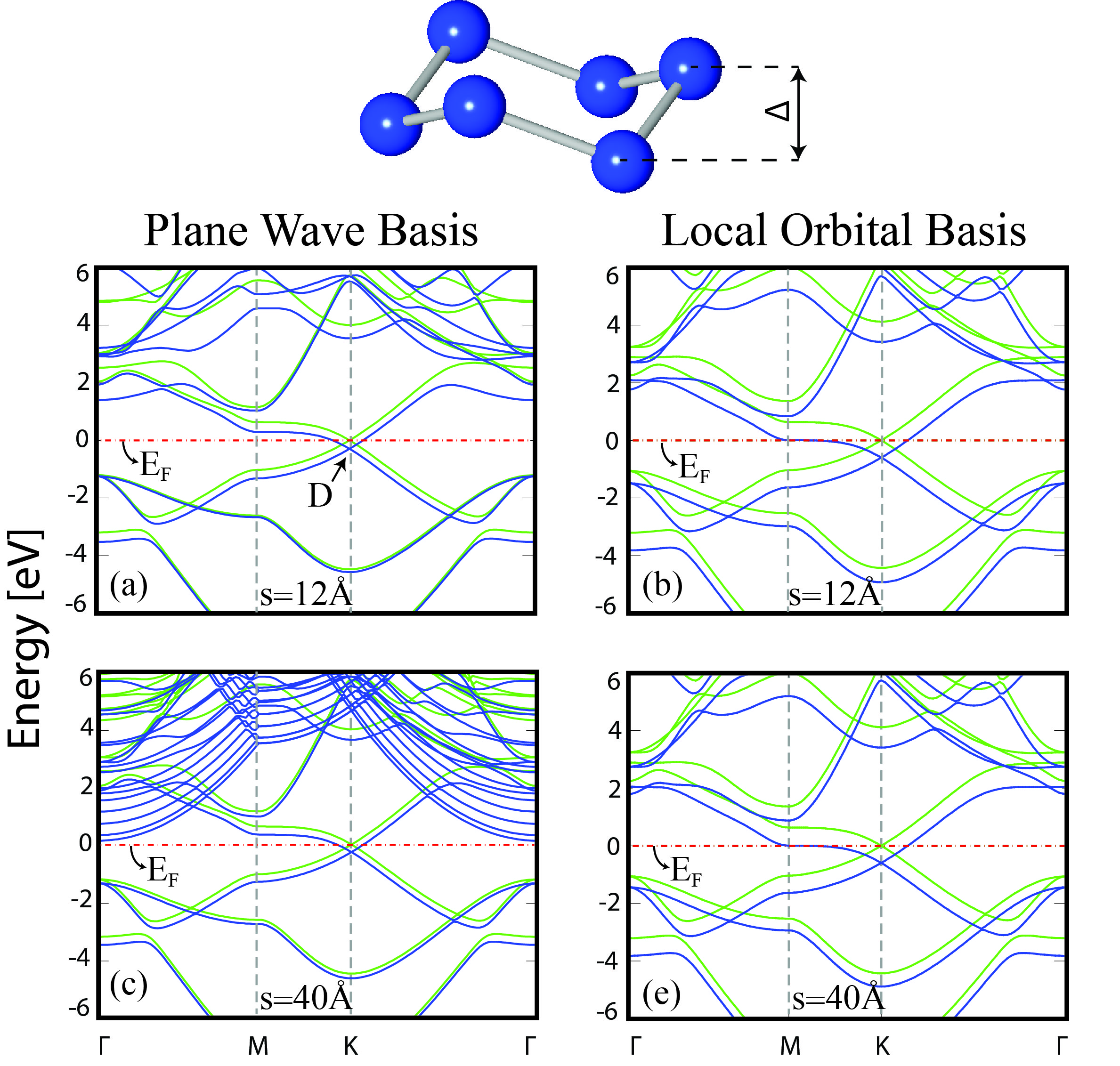}
\caption{(Color Online) Modification of the energy band structure of silicene when charged with $Q=-0.2$ electrons per primitive cell: (a) Energy bands of neutral and charged cases calculated for $s$=12 \AA~ using plane wave (PW) basis set are shown by green(light) and blue(dark) lines, respectively. Dirac points are lowered below the Fermi level upon charging. (b) The same system in (a) is treated using the atomic orbital (AO) basis set. (c) Same as (a) except for $s$=40 \AA. Parabolic bands touching the Fermi level are due to the free electron like states confined to the quantum well centered at $s/2$ as described in Fig.~\ref{fig3} (a). (d) Same as (b) except for $s$=40 \AA. As seen results obtained using AO calculations do not depend on $s$. The inset at the top shows the buckled honeycomb structure of the silicene with buckling $\Delta = 0.45$ \AA.}
\label{fig1}
\end{center}
\end{figure}

The dependence of the electronic structure of silicene when charged or subjected to perpendicular electric field is also of current interest. The stability of this nanostructure in the buckled honeycomb geometry has been proven through ab-initio phonon calculations and extensive high temperature, first-principle molecular dynamics simulations by Cahangirov \textit{et al.}\cite{seymur}. Similar to graphene, the $\pi$ and $\pi^*$ like bands of 2D silicene cross each other linearly at the Fermi level leading to massless Dirac Fermion behavior with a Fermi velocity $v_{F} \sim c/300$, ($c$ being the speed of light) and they exhibit an ambipolar character with perfect electron-hole symmetry\cite{seymur}. Interestingly, quasi one-dimensional armchair nanoribbons of silicene were predicted to be semiconductors and they show a family behavior like graphene\cite{seymur,seymur2}. The synthesis of single-layer silicene\cite{lelay1, lelay2} which was achieved recently, has corroborated theoretical predictions by eliminating doubts about whether such a material can exist even though the parent silicon does not form a layered structure. Like silicene, germanene also displays similar properties. Both silicene and germanene lack strong $\pi$-$\pi$ interaction necessary for planar geometry, but they are stabilized through the $sp^3$-hybridization followed by the $sp^2$-dehybridization\cite{seymur,hasan} leading to buckling.

Normally, electric field applied perpendicularly to the silicene or germanene layers, ($E_{\perp}$), breaks the symmetry between two Si or Ge-layers formed by these atoms situated at the alternating corners of buckled hexagons. Expectantly, linearly crossing bands split to open a band gap. Such a gap opening did not occur in graphene because of the planar geometry of constituent carbon atoms. In fact, the gap opening of linearly crossing bands of silicene under $E_{\perp}$ have been studied earlier by two papers\cite{falko,nano}. One of them\cite{falko} carried out Density Functional (DFT) calculations within periodic boundary conditions (PBC) using plane wave (PW) basis set and reported that the band gap depends on the vacuum spacing $s$ between adjacent layers.

\begin{figure}
\begin{center}
\includegraphics[width=12cm]{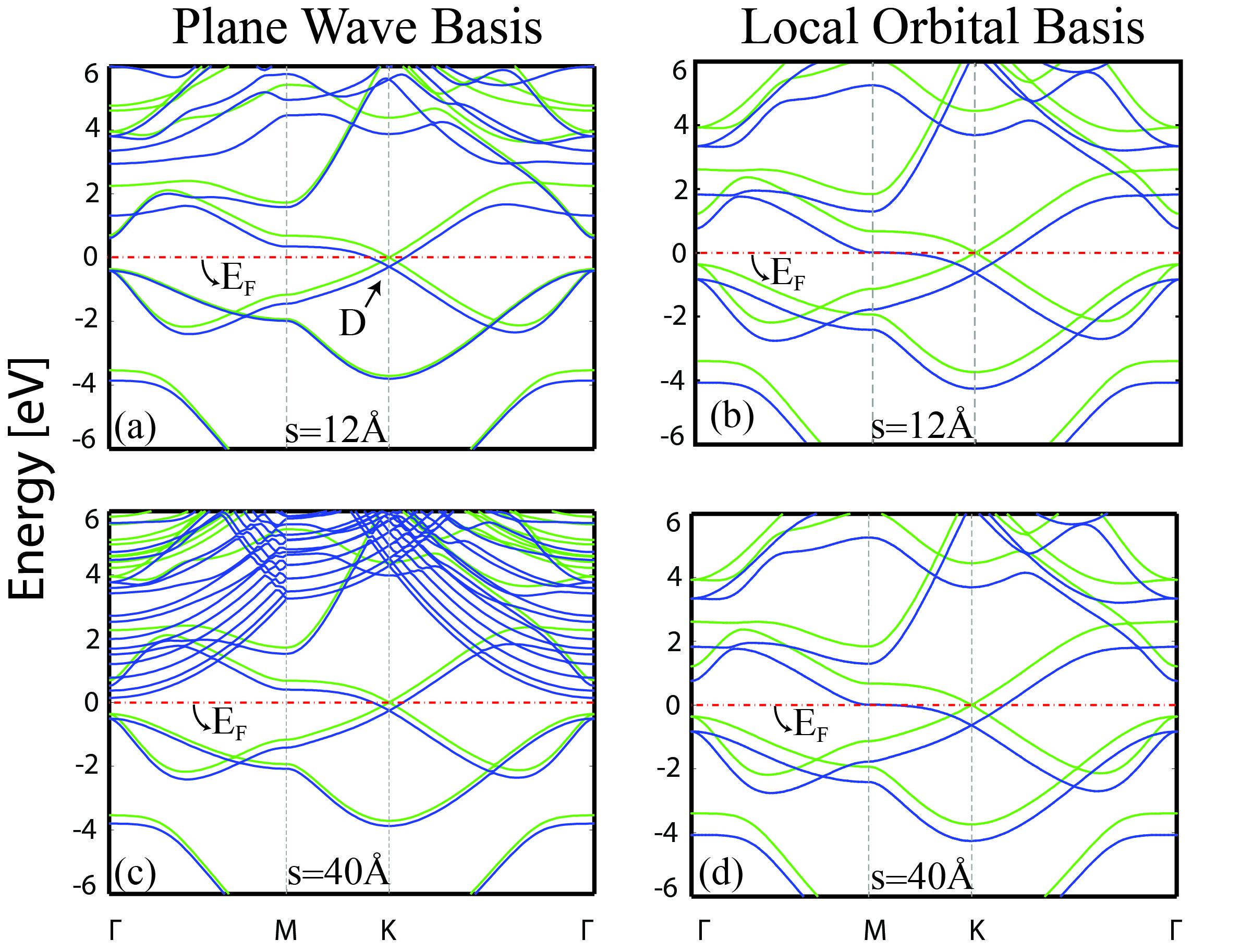}
\caption{(Color Online) Same as Fig.~\ref{fig1} for germanene.}
\label{fig2}
\end{center}
\end{figure}

\section{Methodology}
In this paper, we investigated the effects of charging and applied perpendicular electric field on the electronic, magnetic and chemical properties of silicene and germanene by performing first-principles DFT spin-polarized and spin-unpolarized calculations using PBC within supercell geometry with varying vacuum spacing $s$. We used PW, as well as local, atomic orbital (AO) basis sets. First, we clarified how the effects of charging and perpendicular electric field calculated from first-principle PW method can depend on the  vacuum spacing $s$ as an artifact of the method. Furthermore, we showed that this artifact can be eliminated if silicene or germanene, which are exerted by a $E_{\perp}$ or charged by excess electrons $Q$ are treated using the AO basis set. With the premises that silicene and germanene can attain new functionalities through the charging and the applied electric field, present results will be crucial for future theoretical and experimental studies. In order to show the effects of external effects, such as $E_{\perp}$ and $Q$, we did not consider the spin-orbit coupling, which also gives rise to a small band opening. PW and local AO basis sets are used in numerical calculations through VASP\cite{vasp1} and SIESTA\cite{siesta} packages, respectively.

The exchange-correlation potential is approximated by generalized gradient approximation using Perdew, Burke and Ernzerhof (PBE) functional\cite{pbe}. Dipole corrections\cite{payne} are applied in order to remove spurious dipole interactions between periodic images for the neutral calculations. For all structures studied in this paper, the geometry optimization is performed by the conjugate gradient method by allowing all the atomic positions and lattice constants to relax. The convergence for energy is chosen as 10$^{-5}$ eV between two consecutive steps. In atomic relaxations, the total energy is minimized until the forces on atoms are smaller than 0.04 eV/\AA.

A basis set with kinetic energy cutoff of 500 eV and projector-augmented wave potentials\cite{paw} are used in the PW calculations\cite{vasp1}. In AO calculations\cite{siesta} the eigenstates of the Kohn-Sham Hamiltonian are expressed as linear combinations of numerical atomic orbitals. A 250 Ryd mesh cut-off is chosen and the self-consistent calculations are performed with a mixing rate of 0.2. Core electrons are replaced by norm-conserving, nonlocal Truoiller-Martins pseudopotentials\cite{tmp}. The grid of \textbf{k}-points used is 19x19x1 for AO calculations and 12x12x1 for PW calculations, which are determined by a convergence analysis with respect to the number of grid-points.

\begin{figure}
\begin{center}
\includegraphics[width=12cm]{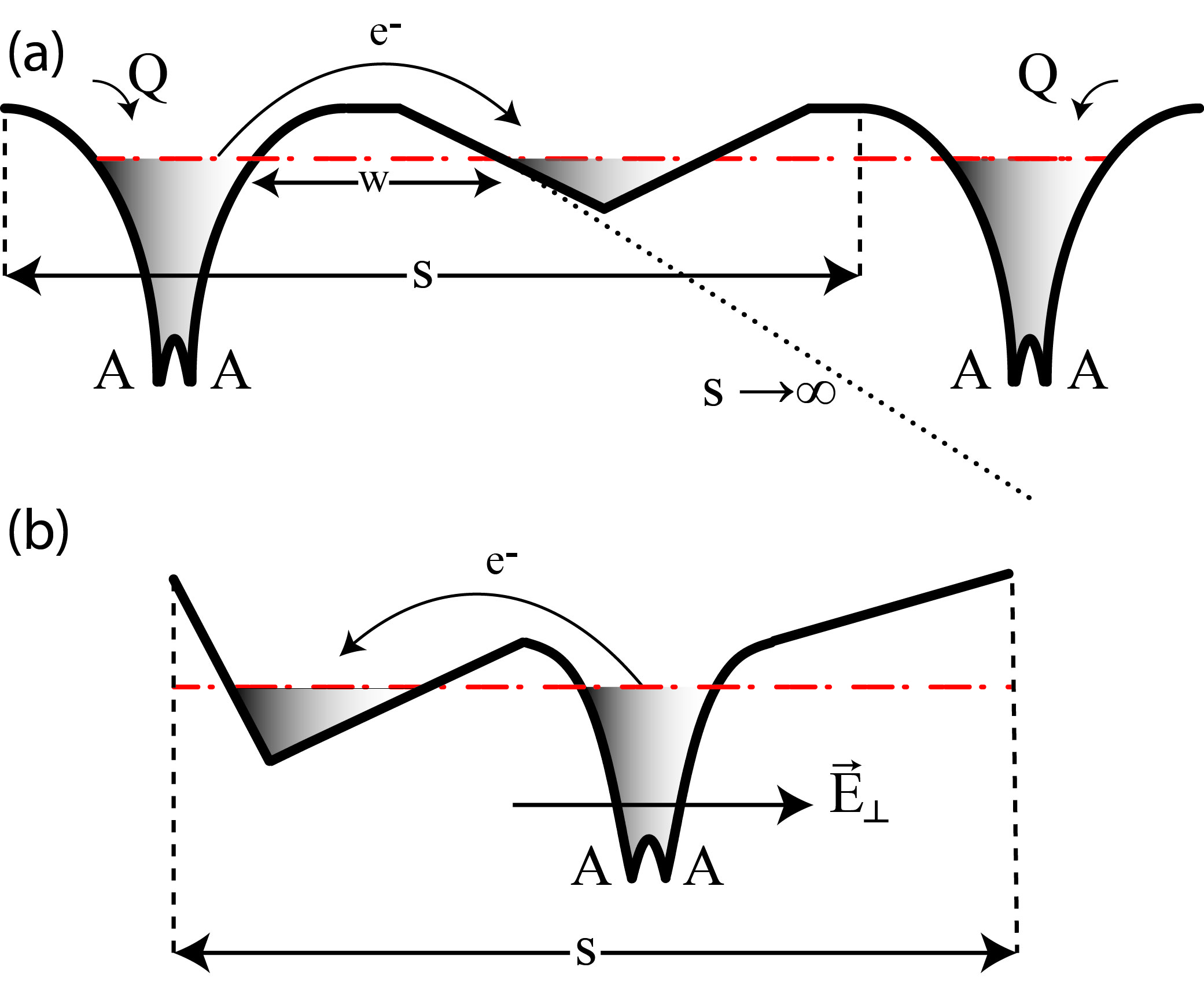}
\caption{(Color Online) (a) A schematic description of plane-averaged electronic potentials $\bar{V}(z)$ explains how the excess electrons of silicene or germanene charged $Q <$ 0 can spill to the vacuum region, when treated using PW within PBC. "A" stands for atomic planes. (b)Same for the electric field $E_{\perp}$, exerting perpendicular to silicene and germanene planes. The vacuum spacing between the layers is denoted by $s$.}
\label{fig3}
\end{center}
\end{figure}

\section{Results}
Before we start to investigate the effects of charging on the electronic structure of silicene and germanene, we first examine the limitations of PBC method, where two-dimensional silicene as well as germanene layers separated by large spacing $s$ are repeated periodically along the $z$-axis. We carried out first-principles PW calculations as well as AO calculations and investigated the effect of charging on silicene and germanene layers. Throughout the paper, $Q <0$ indicates the negative charging, namely the number of excess electrons per primitive cell (corresponding to the surface charge density $\bar{\sigma}= Q/A$ in $Coulomb/m^2$, $A$ being the area of the primitive cell); $Q>0$  indicates the positive charging, namely number of depleted electrons per primitive cell;  and $Q=0$ is the neutral cell.

The calculated electronic structure of silicene and germanene are presented in Fig.~\ref{fig1} and Fig.~\ref{fig2}. For neutral case ($Q$=0), the work function, i.e. the minimum energy that must be given to an electron to release it from the single layer is the difference of the reference vacuum energy and Fermi level. The value of work function extracted from the band structure is $\Phi$=4.57 eV ($\Phi$=4.37 eV for germanene). As seen, the calculated electronic structure using PW and AO are similar. Minute differences originate from different pseudopotentials used in calculations. Moreover, the perfect convergence of different basis sets can be achieved only by using very large cutoff values.

The negative charging can lead marked effects in the electronic structure of single layer silicene and germanene. When charged with $Q=-0.2$ electrons per primitive cell, the energy bands calculated for the vacuum spacing $s=12$ \AA~ using PW in Fig.~\ref{fig1} (a) and AO basis sets in Fig.~\ref{fig1} (b) are similar. The Dirac points dip below the Fermi level and the semimetallic silicene becomes a metal. Similar effect is seen also in the band structure of germanene in Fig.\ref{fig2}. However, when the same charged systems are treated with a relatively larger vacuum spacing of $s=40$ \AA, the PW results dramatically deviate from those of AO.  Parabolic bands are lowered and eventually the Fermi level is pinned as shown in Fig.~\ref{fig1} (c). Such a situation does not occur when AO basis set is used. Hence, results obtained from AO calculations are independent of the size of vacuum spacing as seen in Fig.~\ref{fig1} (d).

The cause of this spurious dependence of vacuum size can be understood through the ($x,y$)-plane averaged self-consistent field (SCF) electronic potential along the $z$-direction. A schematic description of the situation, which explains how the excess electrons for negatively charged silicene or germanene layers can spill into the vacuum region as a result of the above spurious effect is shown in Fig.~\ref{fig3} (a). Earlier we made such analysis by plane-averaging of 3D, SCF one-electron potential of graphene\cite{topsakal1}. Here the same situation is arising; as schematically described, $\bar{V}(z)$ is lowered and passes through a minimum at the middle of vacuum spacing. Normally, for $Q$=0 $\bar{V}(z)$ makes sharp dips at the atomic layers and rises and flattens at $s/2$. The maximum of $\bar{V}(z)$ corresponds to the vacuum potential.  However, in the actual case of single silicene (or germanene) layer, which is charged with $Q <$0, $\bar{V}(z)$ passes through a maximum near the surface and goes to $-\infty$ as $z \rightarrow \pm \infty$  at both sides. Then, in view the 1D WKB approximation, the electrons in the silicene (or germanene) layer might have spilled to vacuum only if they could tunnel across a wide triangular barrier. The width of the barrier, $w$ decreases with increasing negative charging. Under these circumstances, the spilling of electrons must have been negligibly low and hence excess electrons are practically trapped in the layer, if the value of $|Q|$ is not very high. However, when treated within PBC, $\bar{V}(z)$ of periodically repeating silicene (or germanene) layers with a vacuum spacing $s$ between them can make a dip reminiscent of a quantum well at the center of the spacing as shown in Fig.~\ref{fig3} (a). Under these circumstances, Kohn-Sham Hamiltonian using PW can acquire solutions in this quantum well, which are localized along the $z$-direction, but free-electron like in the ($x,y$)-plane parallel to Si planes. These states form the parabolic bands in the ($k_{x},k_{y}$) plane as shown in Fig.~\ref{fig1} (c) and Fig.~\ref{fig2} (c). If the quantum well dips below the Fermi level with increasing $s$ or $Q$, excess electrons in silicene or germanene start to be accommodated in these 2D free-electron like bands. This ends up with the spilling of excess electrons into the vacuum region. The amount of excess electrons spilled to the vacuum spacing increases with increasing $s$ and also increases with increasing negative charging at a fixed value of $s$. The situation in Fig.~\ref{fig1} (c) and Fig.~\ref{fig2} (c), where the Fermi level pinned by the parabolic bands of electrons spilled to the vacuum for large $s$, is an artifact of the PBC and give rise to the dependence on the size of vacuum.

As for AO calculations, the spurious quantum well like structure at the middle of vacuum spacing between silicene or germanene layers in PBC are devoid of basis set and hence cannot support the bound electronic states, since the local atomic orbitals are placed only at the atomic sites. This is why $\bar{V}(z)$ calculated by AO does not accommodate excess electrons in its minimum between silicene layers and hence the parabolic bands seen in Fig.~\ref{fig1} (c) and Fig.~\ref{fig2} (c) do not appear in Fig.~\ref{fig1} (d) and Fig.~\ref{fig2} (d), when it is further lowered below the Fermi level with increasing $s$. The behavior of $\bar{V}(z)$ at the proximity of the surface is similar to the actual case, where $s \rightarrow \infty$. Accordingly, excess negative charge is prevented from spilling into the spacing between adjacent graphene layers, and hence remains in silicene as in the actual case consisting of one single silicene or germanene charged by $Q<$0.

In contrast to the  above shortcomings arising in PW method in treating the negative charging $Q <$0, results obtained by AO and PW are in reasonable agreement for $Q \geq 0$, since $\bar{V}(z)$ passes through a maximum at the center of $s$ between graphene layers. Accordingly, a method using PBC and PW basis set is not affected significantly from the size of vacuum spacing $s$, if $Q \geq 0$. For positively charging, the Fermi level shifts down the Dirac points and the material becomes hole-doped and hence is metallized.

\begin{figure}
\begin{center}
\includegraphics[width=12cm]{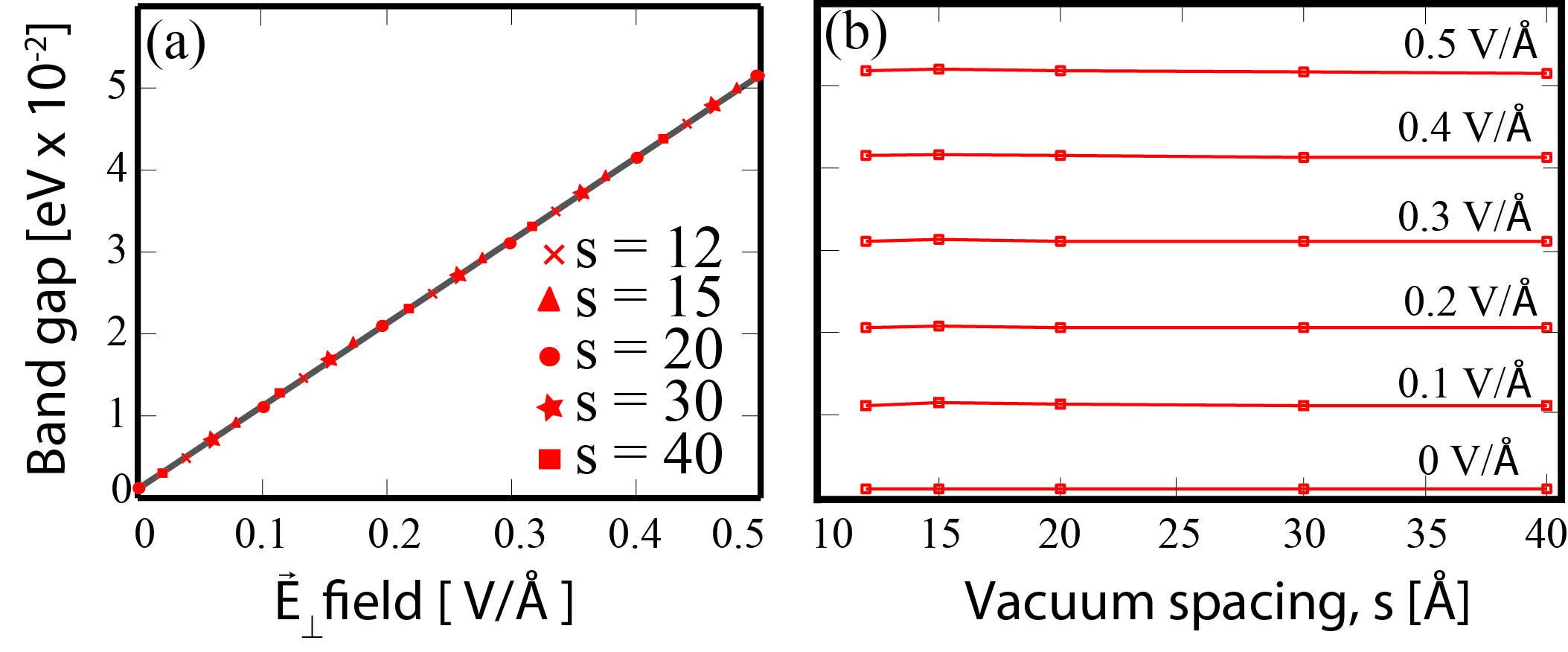}
\caption{(Color Online) (a) Linear variation of band gap opening $\Delta E$, versus applied electric field $E_{\perp}$ for silicene. When obtained using AO, band gap values calculated for various $s$ fall on the same line. (b) Gap opening $\Delta E$ as a function of applied electric field $E_{\perp}$ calculated for different vacuum spacings $s$. Gap openings induced by $E_{\perp}$ does not depend on the vacuum spacing $s$, when calculated using AO.}
\label{fig4}
\end{center}
\end{figure}

In the case of $E_{\perp}$, the mirror symmetry between the top side and the bottom side of silicene is broken, and hence the plane averaged electronic potential exhibits a sawtooth like variation in PBC.  At high $E_{\perp}$  and large $s$, a quantum well like structure occurs at the lowered side of $\bar{V}(z)$. When treated by PW method, this quantum well like structure can dip below the Fermi level and hence electrons of silicene which can be accommodated in this quantum well end up with the electron spilling to vacuum by resulting in a similar situation above for $Q<$0. This situation is schematically described in Fig.~\ref{fig3} (b). This is the reason why the band gap opening under $E_{\perp}$ is depended on $s$ in DFT calculations using the plane wave basis set\cite{falko}. As a matter of fact, in the band structure presented in Fig. 3 (c) of the recent plane wave based study\cite{falko}, at high $E_{\perp}$ one sees a free electron like band touching the Fermi level. However, when treated by AO method, calculated values of the band gap as a function of $E_{\perp}$ fall on the same line as shown in Fig.~\ref{fig4} (a). This is better seen in Fig.~\ref{fig4}(b) demonstrating that band gaps induced in different values of $E_{\perp}$ are practically independent of $s$. The behavior of gap opening with applied $E_{\perp}$ in germanene is similar to that in silicene.

Spin-polarized calculations indicate that the non-magnetic ground state of silicene and germanene is practically  unaltered upon charging or under electric field. However, the permanent magnetic moment of a Ti atom adsorbed to silicene and germanene exhibit significant changes upon charging or under the electric field. However, it is argued that ordinary DFT may not be appropriate to treat the strongly correlated $d$-orbitals, such as in adsorbed Ti atom. Often approaches such as DFT+U are used to address this issue\cite{dudarev}. Surprisingly, in most cases concerning Ti adsorbed on Si and graphene surfaces, DFT  has provided  reasonable predictions. Earlier, we examined the effects of correlation by performing LDA+U calculations\cite{dudarev}. Since U parameters were not available for Ti adsorbed on Si surfaces, we carried out calculations by taking U as a parameter\cite{durgun}. We also performed similar calculations for transition metal dichalcogenides, ScO2, NiO2 and WO2 in single layer honeycomb structure\cite{ataca}. We found that in these studies up to high values of U our predictions obtained by DFT remained valid. Energetically, the most favorable bonding site of the Ti adatom on silicene and germanene is found to be the hollow site, above the center of the hexagon. We investigate the magnetic properties of Ti covered silicene and germanene using supercell geometry in different levels of coverage, since the resulting magnetic ground states depend on the Ti-Ti coupling and hence on the coverage of Ti atom.

First, we consider the case of very weak Ti-Ti coupling by using uniform coverage in which one Ti adatom is adsorbed to each (4 x 4) supercell of silicene(germanene) corresponding to $\Theta$=1/32 (i.e. one Ti atom per 32 Si(Ge) atoms) and leading to the Ti-Ti distance of 15.39\AA (16.06\AA). In the neutral case, both silicene and germanene have a spin-polarized, ferromagnetic (FM) ground state with a magnetic moment of $\mu = 2.44$ $\mu_B /cell$ (i.e per (4 x 4) supercell). The magnetic moment of the spin-polarized state of silicene increases to $\mu = 3.00$ $\mu_B /cell$ ($\mu = 2.55$ $\mu_B /cell$ for germanene) for $Q=$ +1.0 e/cell. For the excess electronic charge of $Q= -1.0 e/cell$ the magnetic moment increases to $\mu = 2.77$ $\mu_B /cell$ ($\mu = 1.44$ $\mu_B /cell$ for germanene). The variation of $\mu$ is due to the accommodation of different electronic charges of Ti $3d$-states for different values of $Q$. Similar effect can be generated also by the static electric field. Spin-polarization of Ti adsorbed silicene and germanene show also significant changes with applied electric field $E_{\perp}$. While their magnetic moments become unaltered for $E_{\perp}= +1.0 V/$\AA~ (i.e. $E_{\perp}$ is directed towards Ti adatom), they increase to $\mu = 3.55 \mu_B /cell$ for $E_{\perp}=-1.0 V/$\AA. Apparently, either by charging of Ti+silicene (Ti+germanene) $(4\times4)$ complex or by exerting an electric field, $E_{\perp} <0$ one can modify the occupations of $3d$-orbitals and hence can change the net magnetic moment.

When a significant Ti-Ti interaction sets in, the magnetic ground states show interesting changes as presented in Table I. We consider the case of significant Ti-Ti coupling by using uniform coverage in which 4 Ti adatoms are adsorbed to each $(4\times4)$ supercell of silicene(germanene) corresponding to $\Theta=1/8$ coverage and leading to a Ti-Ti distance of 7.69\AA~(8.03\AA). This supercell geometry allows us to treat the antiferromagnetic (AFM) order. Similar to the coverage of $\Theta=1/32$, silicene and germanene has FM spin polarized FM magnetic ground states at $\Theta=1/8$, except the case of AFM ground state occurring under $E_{\perp}$=1 V/\AA. Also we predict that an external effect like $Q$ or $E_{\perp}$ causing charge depletion from Ti adatom give rise to an increase in the magnetic moment.

\begin{table}
\caption{Magnetic ground states (AFM: antiferromagnetic or FM: ferromagnetic) and magnetic moments in $\mu_B$  per Ti atom (i.e. per (2x2) cell) calculated as a function of charging and applied perpendicular electric field. Calculations are performed in the (4x4) supercell of silicene and germanene having 4 uniformly adsorbed Ti atoms. $E_{\perp}(+1)$ denotes for positive $E_{\perp}$=1 V/\AA~(directed towards Ti atom) and vice versa for $E_{\perp}(-1)$. $Q(+1e)$ corresponds to the charging where one electron is removed from the (4x4) supercell and vice versa for $Q(-1e)$.}
\label{table1}
\begin{center}
\begin{tabular}{ccccc}
\hline  \hline
& \multicolumn{2}{c}{Silicene + Ti}  & \multicolumn{2}{c}{Germanene + Ti} \\
$E_{\perp}$[V/\AA] or Q[e] & Ground State & Mag. Moment [$\mu_B$] & Ground State & Mag. Moment [$\mu_B$] \\
\hline
$E_{\perp}(+1)$ & AFM & 0.00 & FM& 2.50 \\
$E_{\perp}(-1)$ & FM & 3.11 & FM & 2.39 \\
$Q(+1e)$ & FM & 3.14 & FM & 2.25 \\
$Q(-1e)$ & FM & 2.86 & FM & 2.14 \\
neutral & FM & 3.00 & FM & 2.06 \\
\hline
\hline
\end{tabular}
\end{center}
\end{table}

 We predict that silicene uniformly covered with Ti atom at $\Theta=1/8$ is spin-polarized and has the permanent magnetic moment of $\mu$=3.0 $\mu_B$ and is a half-metal: namely it is metal for one spin direction, but semiconductor for the opposite spin direction. Accordingly, this materials transport electrons only for one spin direction and can function as a spin valve. While silicene has AFM ground state with zero net magnetic moment under $E_{\perp}$=1 V/\AA, the ground state changes to FM when the direction of the electric field is reversed causing charge transfer towards to Ti atom.

\section{Conclusions}

In conclusion, single silicene and germanene can attain useful functionalities by charging and by perpendicular electric field. While charging maintains the symmetry between both sides of honeycomb structure, this symmetry is broken down by the electric field perpendicular to the layer. As a result, linearly crossing bands split, where the band gap becomes linearly dependent on the value of the electric field. Also the work function as well as the binding energies of foreign adatoms become side specific. In particular, the occupations of $3d$-orbitals of transition metal atoms adsorbed to silicene or germanene can be modified by charging or by applied electric field, which in turn, give rise to important changes in magnetic moments. We showed that the vacuum space which can affect the calculated properties of silicene under excess charge or electric field can be eliminated by using local orbital basis sets.

\section{Acknowledgements}
This work was supported by TUBITAK and the Academy of Sciences of Turkey(TUBA). Part of the computational resources has been provided by TUBITAK ULAKBIM, High Performance and Grid Computing Center (TR-Grid e-Infrastructure) and UYBHM at Istanbul Technical University through Grant No: 2-024-2007. Dr. H. H. G\" urel acknowledges the support of TUBITAK-BIDEB.


\section*{References}
\begin{thebibliography}{100}

\bibitem{topsakal1}
Topsakal, M. ; Ciraci, S. Effects of static charging and exfoliation of layered crystals. Phys. Rev. B \textbf{85}, 045121 (2012).

\bibitem{seymur}
Cahangirov, S.; Topsakal, M; Akturk, E.; Sahin, H.; Ciraci S. Two- and One-Dimensional Honeycomb Structures of Silicon and Germanium. Phys. Rev. Lett. \textbf{102}, 236804 (2009).

\bibitem{seymur2}
Cahangirov, S.; Topsakal, M.; Ciraci, S. Armchair nanoribbons of silicon and germanium honeycomb structures. Phys. Rev. B \textbf{81}, 195120 (2010).

\bibitem{hasan}
Sahin, H.; Cahangirov, S.; Topsakal, M.; Bekaraoglu, E.; Akturk, E.; Senger, R.T.; Ciraci, S. Monolayer honeycomb structures of group-IV elements and III-V binary compounds: First-principles calculations. Phys. Rev. B \textbf{80}, 155453 (2009).

\bibitem{lelay1}
Vogt, P; De Padova, P.; Quaresima, C.; Avila, J.; Frantzeskakis, E.; Asensio, M.C.; Resta, A.; Ealet B.; Le Lay,  G. Silicene: Compelling Experimental Evidence for Graphenelike Two-Dimensional Silicon.  Phys. Rev. Lett. \textbf{108}, 155501 (2012)

\bibitem{lelay2}
Aufray, B.; Kara, A.; Vizzini, S.; Oughaddou, H.; Leandri, C.; Ealet, B.; Le Lay, G. Evidence of graphene-like electronic signature in silicene nanoribbons. Appl. Phys. lett. \textbf{96}, 261905 (2010).

\bibitem{falko}
Drummond, N. D.; Z\' olyomi, V.;  Fal'ko, V.I.  Electrically tunable band gap in silicene. Phys. Rev. B \textbf{85}, 075423 (2012).

\bibitem{nano}
Zeyuan, N. ; Qihang, L.; Kechao, T.;  Jiaxin, Z.; Zhou, J.; Rui, Q.; Zhengxiang, G.;  Dapeng Y.; Jing, L. Tunable Bandgap in Silicene and Germanene. Nano Lett. \textbf{12}, 113 (2012).

\bibitem{paw}
Blochl, P.E. Projector augmented-wave method.  Phys. Rev. B \textbf{50}, 17953 (1994).

\bibitem{pbe}
Perdew, J.P.; Burke, K.; Ernzerhof, M.  Generalized Gradient Approximation Made Simple. Phys. Rev. Lett. \textbf{77}, 3865 (1996).

\bibitem{vasp1}
Kresse, G.; Furthmuller, J. Efficient iterative schemes for ab initio total-energy calculations using a plane-wave basis set. Phys. Rev. B \textbf{54}, 11169 (1996).

\bibitem{siesta}
Soler, J. M.  et al.  The SIESTA method for ab initio order-N materials simulation.  J. Phys.: Condens. Matter \textbf{14}, 2745 (2002).

\bibitem{payne}
Markov, G.; Payne,  M.C. Periodic boundary conditions in ab initio calculations. Phys. Rev. B \textbf{51}, 4014 (1995).

\bibitem{tmp}
Troullier, N.; Martins, J.L. Efficient pseudopotentials for plane-wave calculations Phys. Rev. B \textbf{43}, 1993 (1991).

\bibitem{dudarev}
Dudarev, S.L.; Botton, G.; Savrasov,  S.; Humpreys, C.; Sutton,A. Electron-energy-loss spectra and the structural stability of nickel oxide: An LSDA+U study. Phys. Rev. B \textbf{57},1505 (1998).

\bibitem{durgun}
Durgun, E.; Cakir, D.; Akman, N.; Ciraci, S. Half-Metallic Silicon Nanowires: First-Principles Calculations. Phys. Rev. Lett. \textbf{99}, 256806 (2007).

\bibitem{ataca}
Ataca, C.; Sahin, H.; Ciraci, S. Stable, Single-Layer $MX_2$ Transition-Metal Oxides and Dichalcogenides in a Honeycomb-Like Structure J. Phys. Chem. C, 8983-8999 (2011).


\end{thebibliography}
\end{document}